\documentclass[preprint,12pt]{elsarticle}

\usepackage{url}

\usepackage{CJKutf8}

\usepackage{amssymb}
\usepackage{amsmath}

\journal{Journal of Luminescence}

\begin{document}

\begin{frontmatter}



\title{Infrared Spectroradiometry of Sodium Benzoate \\
from 21 to 235\,THz} 


\author[a]{Yoshitaka Okuyama}
\author[a]{Youichi Ishikawa}
\author[a]{Daishi Fujita}

\affiliation[a]{organization={Institute for Integrated Cell-Material Sciences (WPI-iCeMS), Kyoto University},
            addressline={Yoshida}, 
            city={Sakyo-ku},
            postcode={606-8501}, 
            state={Kyoto},
            country={Japan}}

\begin{abstract}
This paper presents an extensive survey of the thermal radiation properties of sodium benzoate. We heated the sample from 313 to 553 K, just below the melting point, while performing an infrared spectroradiometry with an FT-IR spectrometer from 21 to 235 THz (700-7800\,cm$^{-1}$). We have provided a detailed analysis of the infrared spectrum data and a comparison of the absorption spectrum of the same sample. It turned out that the recorded spectra are not only different from ordinary absorption spectra but also carry substantial information about the temperature dependence of the population of vibrationally excited states. We conclude by proposing a hypothesis on the thermal excitation mechanism of vibrational energy levels of molecules consistent with the distinct characteristics of the obtained infrared emission spectra.
\end{abstract}

\begin{keyword}
Fourier-Transform infrared spectroscopy
\sep 
spectroradiometry
\sep 
terahertz
\sep
vibrational spectroscopy
\sep 
infrared emission spectroscopy
\sep 
thermal radiation
\sep 
spontaneous emission




\end{keyword}

\end{frontmatter}



\section{Introduction}
\label{sec:Introduction}

Spectroradiometry, a branch of radiometry that measures the spectral power distribution of a light source across different wavelengths, has played a pivotal role in scientific development \cite{encyclopedia}. Particularly, it is an indispensable tool in astrophysics, as it can determine the surface color temperatures of celestial bodies along with the compositions of their atmospheres from the dark absorption lines.
Despite various derivatives of spectroradiometry, such as flame spectroscopy and plasma atomic emission spectroscopy, most measurements are in the ultraviolet and visible domains. Less attention has been paid to the infrared, except in the steel industry, to examine the deterioration of metal coatings at high temperatures, where thermal radiation gets so intense that the usual attenuated total reflection turns erroneous \cite{Suetaka}. 

We now provide a few comments on the origin of infrared thermal radiation from molecules. Assume the energy distribution of the canonical ensemble of statistical mechanics, that is, a simple working model of thermal excitation. Then, the population of the vibrationally excited state of a molecule $N_{\text{ex}}$ relative to that of the vibrationally ground state $N_{\text{gr}}$ is given by
\begin{align}\label{eq:statmch_estimate}
\frac{N_{\text{ex}}}{N_{\text{gr}}} = e^{-\frac{E_{\text{ex}}-E_{\text{gr}}}{k_{\text{B}}T}}\ ,
\end{align}
where $k_{\text{B}}$ and $T$ are the Boltzmann constant and the temperature of the system, respectively. And $E_{\text{ex}}$ is the vibrational energy of the excited molecule, while $E_{\text{gr}}$ is the ground state energy of the molecule. This formula tells us that the relative population of the vibrationally excited state at 1000\,cm$^{-1}$ over the ground state at 300 K is about $0.007$, and $0.05$ at 500 K. Although vibrationally excited molecules may appear negligible, the total number of molecules collected in palm-sized pieces is so large that one can examine spontaneous emissions from them (see e.g., \cite{Sakurai}) by performing spectroradiometry in the infrared. 

Despite a long history of infrared spectroradiometry, primarily in the steel industry (see e.g., \cite{Frank91,Chiang83,Tobin87-2,Tobin87-1,Lauer84,Gratton:78,LAUER1979395}), most of the research focuses only on the mid-Infrared (mid-IR) region of the spectrum (400-4000\,cm$^{-1}$), or relatively high temperatures over 400 K. In this paper, we take a step forward and perform infrared spectroradiometry of an organic molecule, sodium benzoate, from 21 to 235 THz (700-7800\,cm$^{-1}$), encompassing both mid-IR and near-infrared (NIR) regions over 4000\,cm$^{-1}$. 
For a detailed investigation of the temperature dependence of the infrared (IR) emission of sodium benzoate, we systematically recorded its IR emission from 313 to 553 K in 40-degree increments. This focus helps to elucidate the temperature-dependent spectral features within the studied frequency range.

\section{Experimental procedure}

\begin{figure}[ht]
\centering
\includegraphics[scale=0.4,angle=270]{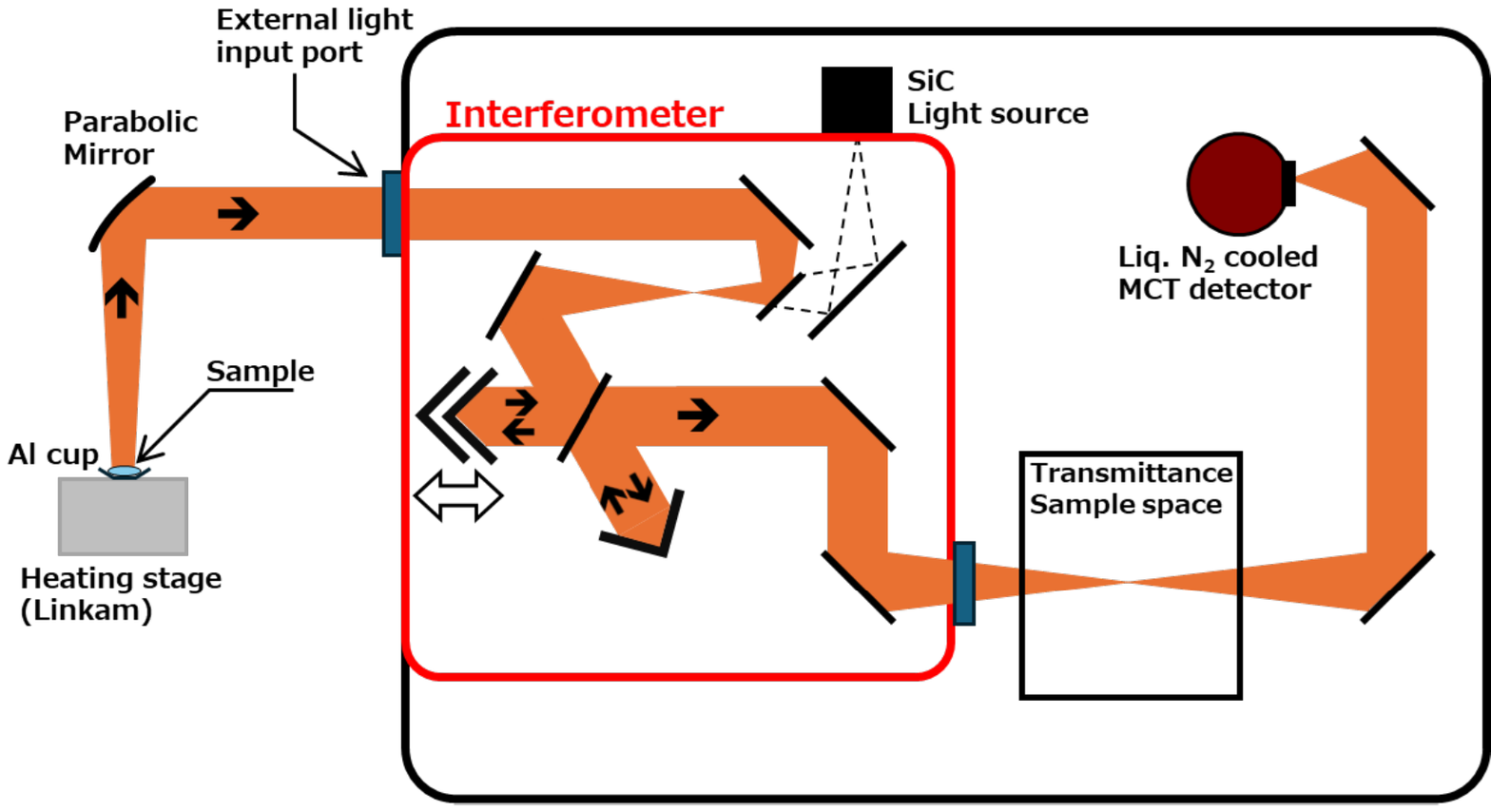}
\caption{A schematic drawing of the infrared emission measurement system.}
\label{fig:Spectrometer}
\end{figure}

\subsection{Experimental setup}

Sodium benzoate was purchased from BLD Pharm (Shanghai, China; catalog number BD136438-25g, lot number BGW609) under the product name lithium benzoate (stated purity 99\%) and was used as received. During a related study, the elemental composition of this lot was examined by inductively coupled plasma atomic emission spectroscopy (ICP-AES) using an ICPE-9820 spectrometer (Shimadzu Corp.). The analysis showed Na = 15.9 wt\% and Li below the detection limit, identifying the material unambiguously as sodium benzoate. All measurements reported in this work were performed on this material. The sample, a powder with a purity of 99\%, was placed in a 10 mm diameter aluminum cup, forming a layer approximately 0.5 mm thick.

 Fig.\,\ref{fig:Spectrometer} shows a schematic of the IR emission measurement system in this paper. IR emission from the sample was directed into the interferometer using an off-axis parabolic mirror with a long focal length. We recorded the IR emission spectra of the powder samples with a JASCO FT/IR-8X Fourier transform infrared spectrometer equipped with a mercury-cadmium-telluride (MCT) detector, cooled to liquid nitrogen temperature, with a spectral resolution of 4.0 cm$^{-1}$ and 90 accumulations. The spectra were obtained while tuning the sample temperature with a Linkam Scientific Instruments stage from 313 to 553 K in 40-degree increments. To ensure reproducibility, we collected six sets of data for every measurement with fixed conditions and used their averages for our analysis. 

\subsection{Method of analysis}
One of the difficulties in spectroradiometry of IR emission lies in the fact that every substance, including the spectrometer and the sample holder, radiates. That is, it is virtually impossible to collect radiation purely from the sample while isolating it from the rest. Hence, we have taken the following measures: Record (single-beam) IR emission spectra of the sample holder, a 10 mm diameter aluminum cup, with and without the sample. Focus on the ratio of their spectra, that is, the relative emission spectra of the sample:
\begin{align}
  \text{Relative IR emission spectrum}=  \frac{\text{IR emission with the sample}}{\text{IR emission without the sample}}\ ,
\end{align}
to extract the IR emission data solely from the specific sample as much as possible while minimizing the effects of background noise and environmental-dependent factors, such as IR emissions from the environment and varying factors relevant to the quantum efficiency of the MCT detector.\footnote{Another factor to be considered is backward IR emissions from the interior optical components of the spectrometer. They give rise to negative contributions to the interferogram with respect to forward emissions from the sample and the environment, making the Fourier-transformed (single-beam) IR emission spectra erroneous, particularly when the sample is around room temperature \cite{Xiao19}. Since we have taken the IR emission spectra of the sample heated over 40${}^\circ$C, well above the room temperature around 24${}^\circ$C, we have assumed that the backward IR emissions only give negligible contributions to our analysis throughout this paper.}

\section{Result and Discussion}

\subsection{Overview of obtained spectra}
\begin{figure}[ht]
\centering
\includegraphics[scale=0.45,angle=270]{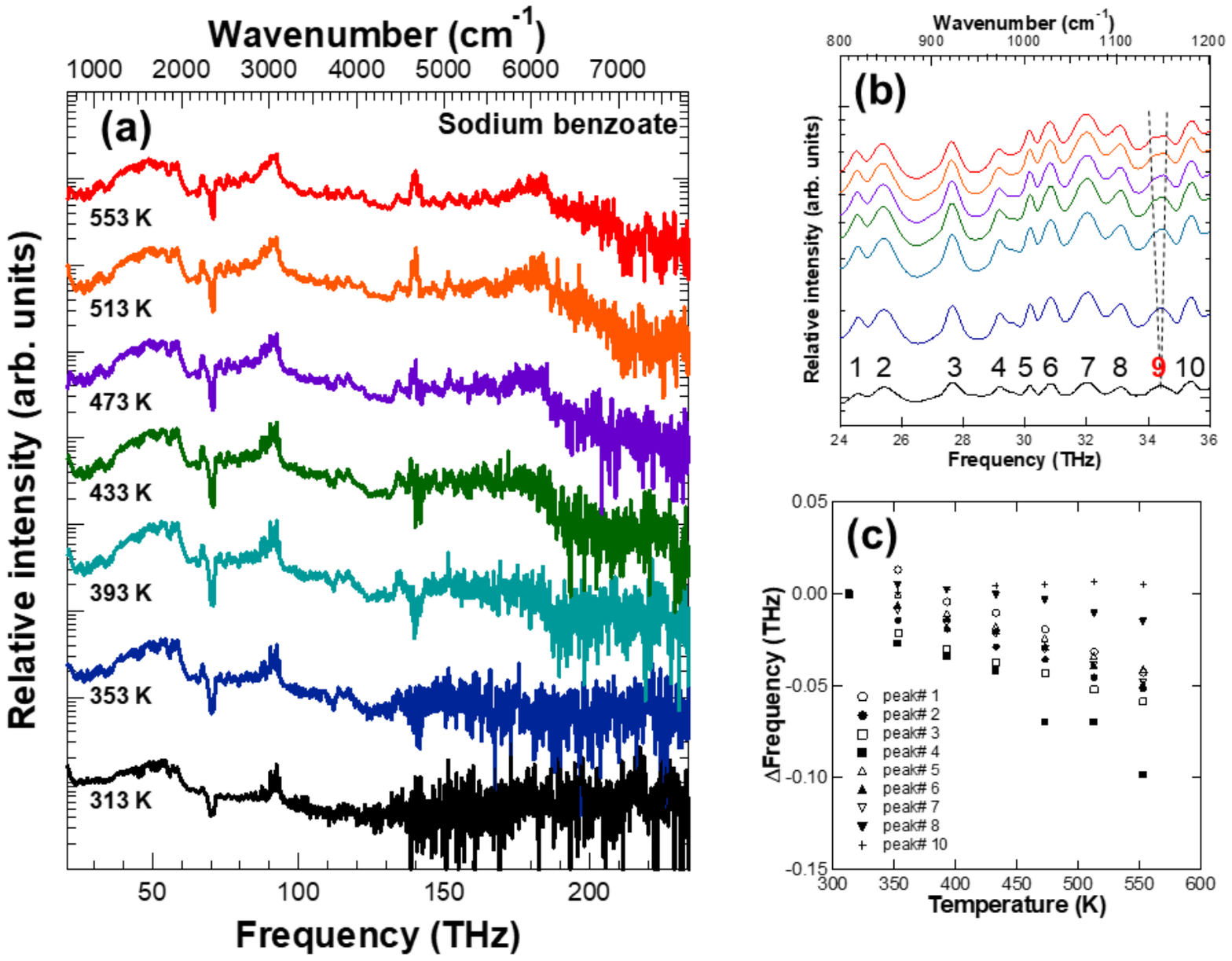}
\caption{(a) Temperature dependence of the relative IR emission spectra of sodium benzoate powder. (b) An enlargement of (a) into the region 24-36 THz (800-1190 cm$^{-1}$). (c) The degree of shifts of all numbered peaks in (b) relative to those at the lowest temperature.}
\label{fig:relative_emission}
\end{figure}

Fig.\,\ref{fig:relative_emission}-(a) shows the relative IR emission spectra of sodium benzoate. The overall spectral profiles in the mid-IR region exhibit little change with increasing temperature. In contrast, the NIR spectra become sharper at elevated temperatures. As an illustration, the spectra around 4600 cm$^{-1}$ appear fuzzy at low temperatures, due to atmospheric CO$_2$ absorption, but become clearer as the sample is heated. This trend suggests that enhanced molecular motion at higher temperatures increases the population of higher vibrational levels, leading to the emergence of more distinct NIR spectral features. 

The assignment of NIR peaks to specific functional groups, following a conventional method \cite{Ozaki}, is as follows. Multiple peaks around 4600 cm$^{-1}$ correspond to the combination bands of the aromatic C-H stretch (3000–3100 cm$^{-1}$), and the C-C stretch of the aromatic ring or the C=O stretch (1400–1600 cm$^{-1}$). The jagged peaks near 5200 cm$^{-1}$ likely correspond to the second overtones or the combination bands of the C-C stretch of the aromatic ring and the C=O stretch (1400–1600 cm$^{-1}$). The peaks around 6000 cm$^{-1}$ can be associated with the first overtone of the aromatic C-H stretch (3000–3100 cm$^{-1}$).

\begin{figure}[ht]
\centering
\includegraphics[scale=0.2]{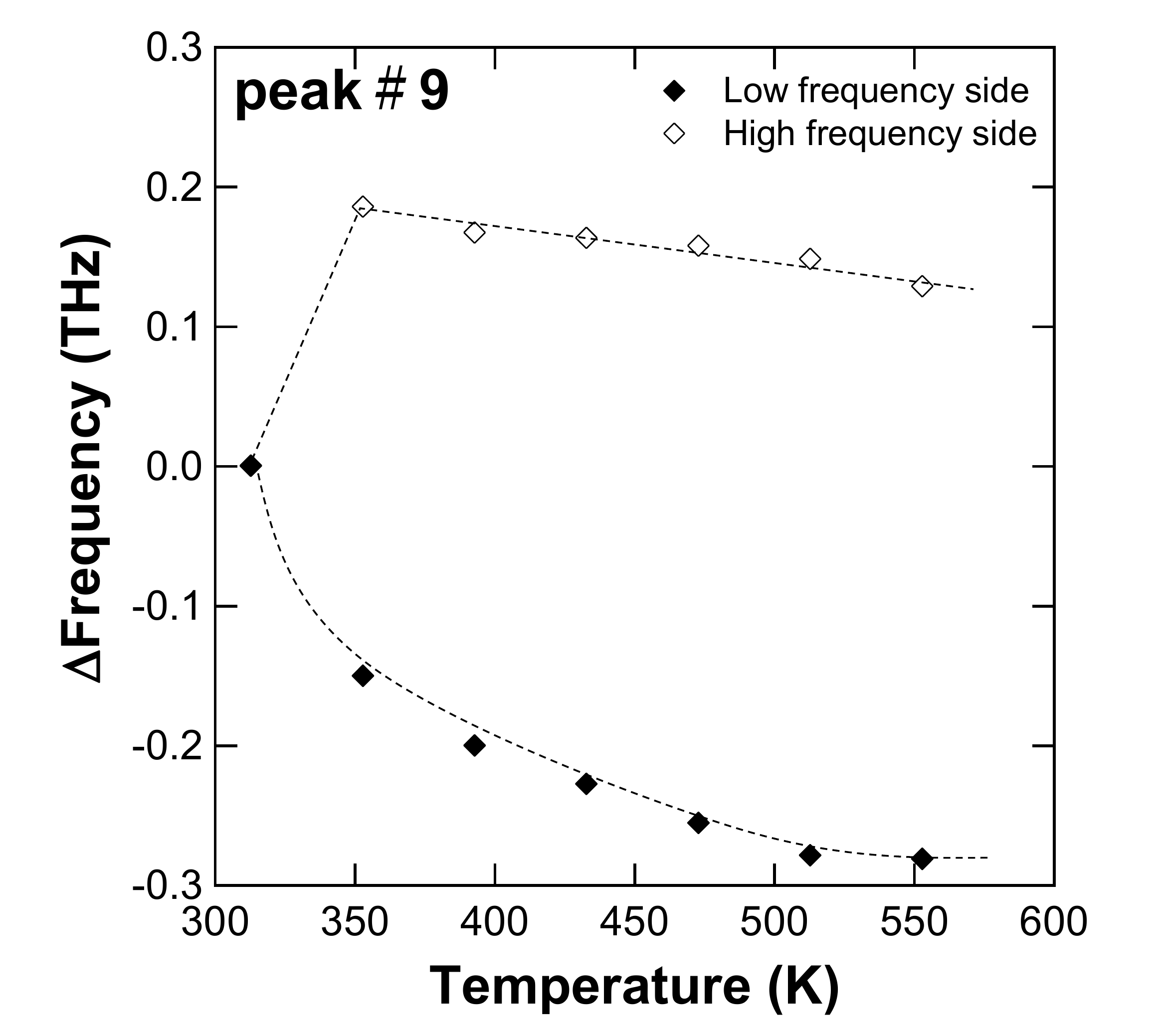}
\caption{The amount of shift of the ninth peak in Fig.\ref{fig:relative_emission}-(b) relative to that at the lowest temperature.}
\label{fig:9th_peak}
\end{figure}

\subsection{Details of mid-IR spectra}
We now provide a more detailed analysis of the mid-IR region in Fig.\,\ref{fig:relative_emission}-(a). For this purpose, we expand in Fig.\,\ref{fig:relative_emission}-(b) a portion of the fingerprint region of the spectra (24-36 THz or 800-1190 cm$^{-1}$), while putting numbers on all peaks.

Figure \ref{fig:relative_emission}-(c) shows the peak shifts relative to those in the lowest temperature, excluding the ninth peak. In general, all peaks broaden as the temperature increases. One of the causes of line broadening may be the shortening of the lifetime of the vibrationally excited states due to increased collisional broadening at higher temperatures. The first eight peaks are red shifted at higher temperatures. Remarkably, the amount of shift is only about 0.1\% of the total wavenumber, which may be due to the robustness of our sample, sodium benzoate, with a high melting point. Moreover, the tenth peak is almost invariant, although it may be due to the strong adjacent band. The ninth peak gradually splits into two pieces, as shown in Fig. \ref{fig:relative_emission}-(b), but both exhibit small redshifts at higher temperatures (see Fig.\,\ref{fig:9th_peak}). 

\subsection{Comparison between mid-IR absorption and emission spectra}

We also recorded the mid-IR absorption spectrum of sodium benzoate at 300 K to compare it with the emission spectrum at 313 K. The infrared absorption spectrum was recorded using an FT-IR spectrometer (FT/IR-4200, JASCO) equipped with a DLATGS (Deuterated L-Alanine Triglycine Sulfate) detector, with 90 accumulations at a resolution of 4.0 cm$^{-1}$. We have prepared the sample as a KBr pellet containing 1.0 wt\% of the analyte. Potassium bromide (KBr, 99\%, CAS No. 7758-02-3) was purchased from Tokyo Chemical Industry (TCI). 

\begin{figure}[ht]
\centering
\includegraphics[scale=0.45,angle=270]{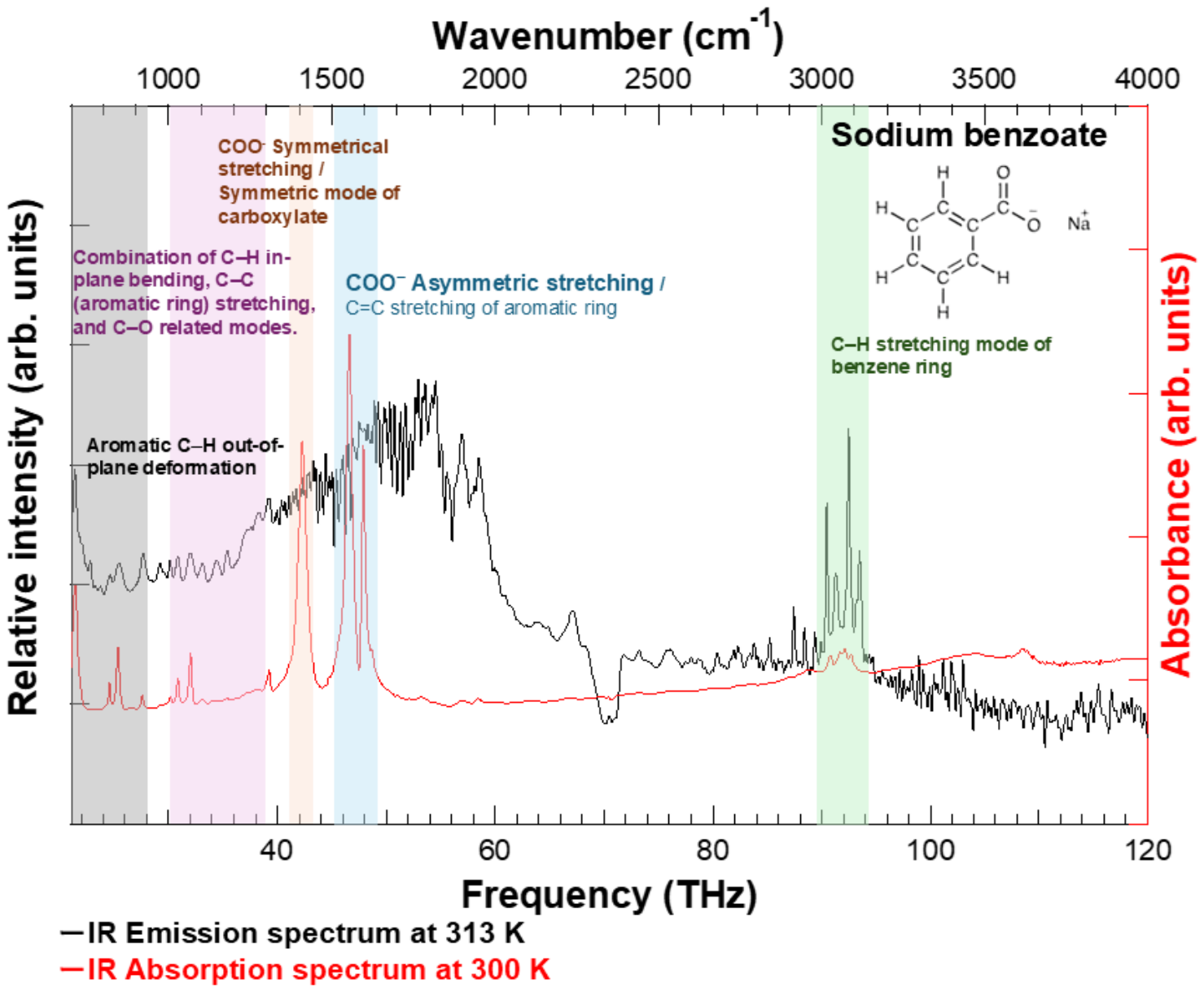}
\caption{Emission and absorption spectrum of sodium benzoate. The black line corresponds to the emission spectrum at 313 K, while the red line represents the absorption spectrum at 300 K.}
\label{fig:emission_vs_asborption}
\end{figure}

Fig.\,\ref{fig:emission_vs_asborption} shows the absorption and emission spectra with a detailed peak assignment according to \cite{REGULSKA2005353}. Notice that the structure of the emission spectrum is distinct from and more involved than the absorption spectrum. More specifically, there are more peaks in the emission spectrum than in the absorption spectrum. They have several peaks in common, but with different intensities. 

We propose that this distinction is responsible for the different mechanisms of absorption and emission measurements. In the former, we monitor the attenuation of light by the sample, during which process molecules in the sample are excited by the absorption of light quanta. Even at low temperatures, some molecules should be in their vibrationally excited states, as we demonstrated in Fig.\,\ref{fig:relative_emission}-(a). However, they are averaged out to some extent in ordinary mid-IR and NIR absorption measurements at room temperature and below, resulting in reasonable agreement with quantum-mechanical calculations of vibrational frequencies at absolute zero. However, this mechanism may not apply to absorption spectroscopy in the THz region, where considerable amounts of molecules should be vibrationally excited even at room temperature, according to a rough estimate from \eqref{eq:statmch_estimate}.

In emission measurements, we detect weak spontaneous emissions from molecules that are vibrationally excited by heat. To provide a consistent explanation of this phenomenon, we should first clarify the mechanism of the thermal excitation of molecules. For this purpose, we would like to use the following analogy to describe our hypothesis. Imagine a staircase. Each step represents a vibrational energy level of a molecule. At the very bottom (ground state) is a hose, from which water flows at a rate corresponding to the temperature, rushing up the stairs. We can analogize the way in which a portion of water moves up and down the stairs due to collisions with others to vibrational excitations and de-excitations as a consequence of inelastic collisions between molecules. Gravity pulls the water downward toward the ground state, but after some time, the amount of water at each step will become stationary. 

Recall that any two excited states of a molecule with nonzero transition dipole moments can undergo spontaneous emission from the higher to the lower. Considering this within our aforementioned analogy, the rate of descent while emitting light quanta is determined from the principle of quantum mechanics, occurring in any pair of steps proportionally to the transition dipole moments of the two states. For instance, spontaneous emissions should occur not only from direct de-excitations to the ground state but also from overtones and combination bands to the fundamentals. In this manner, there should be a cascade-like mechanism and many pathways for spontaneous emission of radiation, resulting in a more structured spectrum with more peaks than in absorption measurements. The split of the ninth peak in Fig.\,\ref{fig:relative_emission}-(b) may be linked to the opening of new channels of spontaneous emission from higher excited states to lower ones at higher temperature.

\section{Conclusion}

In this paper, we performed the infrared spectroradiometry of sodium benzoate from 21 to 235 THz (700-7800\,cm$^{-1}$) at various temperatures. The obtained spectra show strong temperature dependence in the NIR region. In particular, the NIR spectra become sharper as the sample is heated, implying an increase in the population of higher vibrationally excited states. 

In differentiating the absorption and emission spectra of sodium benzoate, we demonstrated that the emission spectrum provides more information than the absorption spectrum. To explain this characteristic feature of the emission spectra, we proposed the cascade-like thermal excitation and spontaneous emission mechanism. However, the complicated structure of the thermal emission prevented us from identifying the origin of all peaks in the emission spectra. More extensive investigations are needed to confirm our hypothesis and to extract much more information from emission spectra.

\paragraph{Acknowledgement}
The authors thank S. Norimoto and S. Matsumoto for their support of our experiment.
This work was supported in part by AMED (JP22am0401020) by CAO, JST FOREST Program (JPMJFR203R), JST‑Mirai (JPMJMI22H5), JSPS KAKENHI (JP23K26777) by MEXT, JKA (KEIRIN RACE promotion funds), Inamori Foundation (InaRIS Fellowship) and Asian Young Scientist Fellowship.








\bibliographystyle{elsarticle-num-names} 
\bibliography{IRemission}

\begin{thebibliography}{13}
\expandafter\ifx\csname natexlab\endcsname\relax\def\natexlab#1{#1}\fi
\providecommand{\url}[1]{\texttt{#1}}
\providecommand{\href}[2]{#2}
\providecommand{\path}[1]{#1}
\providecommand{\DOIprefix}{doi:}
\providecommand{\ArXivprefix}{arXiv:}
\providecommand{\URLprefix}{URL: }
\providecommand{\Pubmedprefix}{pmid:}
\providecommand{\doi}[1]{\href{http://dx.doi.org/#1}{\path{#1}}}
\providecommand{\Pubmed}[1]{\href{pmid:#1}{\path{#1}}}
\providecommand{\bibinfo}[2]{#2}
\ifx\xfnm\relax \def\xfnm[#1]{\unskip,\space#1}\fi
\bibitem[{Worsfold et~al.(2019)Worsfold, Townshend, Poole, and Mir{\'o}}]{encyclopedia}
\bibinfo{author}{P.~Worsfold}, \bibinfo{author}{A.~Townshend}, \bibinfo{author}{C.~F. Poole}, \bibinfo{author}{M.~Mir{\'o}}, \bibinfo{title}{Encyclopedia of Analytical Science}, \bibinfo{publisher}{Elsevier Science}, \bibinfo{year}{2019}.
\bibitem[{Su{\"e}taka and Yates(1995)}]{Suetaka}
\bibinfo{author}{W.~Su{\"e}taka}, \bibinfo{author}{J.~T. Yates}, \bibinfo{title}{Surface Infrared and Raman Spectroscopy: Methods and Applications}, \bibinfo{publisher}{Springer}, \bibinfo{year}{1995}. \DOIprefix\doi{https://doi.org/10.1007/978-1-4899-0942-8}.
\bibitem[{Sakurai(1967)}]{Sakurai}
\bibinfo{author}{J.~J. Sakurai}, \bibinfo{title}{Advanced Quantum Mechanics}, \bibinfo{publisher}{Pearson Education, Incorporated}, \bibinfo{year}{1967}.
\bibitem[{Deblase and Compton(1991)}]{Frank91}
\bibinfo{author}{F.~J. Deblase}, \bibinfo{author}{S.~Compton},
\newblock \bibinfo{title}{Infrared emission spectroscopy: A theoretical and experimental review},
\newblock \bibinfo{journal}{Applied Spectroscopy} \bibinfo{volume}{45} (\bibinfo{year}{1991}) \bibinfo{pages}{611--618}. \URLprefix \url{https://doi.org/10.1366/0003702914337029}. \DOIprefix\doi{10.1366/0003702914337029}. \href{http://arxiv.org/abs/https://doi.org/10.1366/0003702914337029}{{\tt arXiv:https://doi.org/10.1366/0003702914337029}}.
\bibitem[{Chiang et~al.(1983)Chiang, Tobin, and Richards}]{Chiang83}
\bibinfo{author}{S.~Chiang}, \bibinfo{author}{R.~G. Tobin}, \bibinfo{author}{P.~L. Richards},
\newblock \bibinfo{title}{Infrared emission spectroscopy of {CO} on {Ni}},
\newblock \bibinfo{journal}{Journal of Electron Spectroscopy and Related Phenomena} \bibinfo{volume}{29} (\bibinfo{year}{1983}) \bibinfo{pages}{113--118}. \URLprefix \url{https://www.sciencedirect.com/science/article/pii/0368204883800498}. \DOIprefix\doi{https://doi.org/10.1016/0368-2048(83)80049-8}.
\bibitem[{Tobin et~al.(1987)Tobin, Phelps, and Richards}]{Tobin87-2}
\bibinfo{author}{R.~G. Tobin}, \bibinfo{author}{R.~B. Phelps}, \bibinfo{author}{P.~L. Richards},
\newblock \bibinfo{title}{An infrared emission study of the {C=O} stretch vibration of bridge-bonded co on pt(111)},
\newblock \bibinfo{journal}{Surface Science} \bibinfo{volume}{183} (\bibinfo{year}{1987}) \bibinfo{pages}{427--437}. \URLprefix \url{https://www.sciencedirect.com/science/article/pii/S0039602887802194}. \DOIprefix\doi{https://doi.org/10.1016/S0039-6028(87)80219-4}.
\bibitem[{Tobin and Richards(1987)}]{Tobin87-1}
\bibinfo{author}{R.~G. Tobin}, \bibinfo{author}{P.~L. Richards},
\newblock \bibinfo{title}{An infrared emission study of the molecule-substrate mode of {CO} on {P}t(111)},
\newblock \bibinfo{journal}{Surface Science} \bibinfo{volume}{179} (\bibinfo{year}{1987}) \bibinfo{pages}{387--403}. \URLprefix \url{https://www.sciencedirect.com/science/article/pii/0039602887900653}. \DOIprefix\doi{https://doi.org/10.1016/0039-6028(87)90065-3}.
\bibitem[{Lauer and Vogel(1984)}]{Lauer84}
\bibinfo{author}{J.~L. Lauer}, \bibinfo{author}{P.~Vogel},
\newblock \bibinfo{title}{Emission ftir analyses of thin microscopic patches of jet fuel residues deposited on heated metal surfaces},
\newblock \bibinfo{journal}{Applications of Surface Science} \bibinfo{volume}{18} (\bibinfo{year}{1984}) \bibinfo{pages}{182--206}. \URLprefix \url{https://www.sciencedirect.com/science/article/pii/0378596384900448}. \DOIprefix\doi{https://doi.org/10.1016/0378-5963(84)90044-8}.
\bibitem[{Gratton et~al.(1978)Gratton, Paglia, Scattaglia, and Cavallini}]{Gratton:78}
\bibinfo{author}{L.~M. Gratton}, \bibinfo{author}{S.~Paglia}, \bibinfo{author}{F.~Scattaglia}, \bibinfo{author}{M.~Cavallini},
\newblock \bibinfo{title}{Infrared emission spectroscopy applied to the oxidation of molybdenum},
\newblock \bibinfo{journal}{Appl. Spectrosc.} \bibinfo{volume}{32} (\bibinfo{year}{1978}) \bibinfo{pages}{310--316}. \URLprefix \url{https://opg.optica.org/as/abstract.cfm?URI=as-32-3-310}.
\bibitem[{Lauer and King(1979)}]{LAUER1979395}
\bibinfo{author}{J.~L. Lauer}, \bibinfo{author}{V.~W. King},
\newblock \bibinfo{title}{Fourier emission infrared microspectrophotometer for surface analysis. {I} - {A}pplication to lubrication problems},
\newblock \bibinfo{journal}{Infrared Physics} \bibinfo{volume}{19} (\bibinfo{year}{1979}) \bibinfo{pages}{395--412}. \URLprefix \url{https://www.sciencedirect.com/science/article/pii/0020089179900514}. \DOIprefix\doi{https://doi.org/10.1016/0020-0891(79)90051-4}.
\bibitem[{Xiao et~al.(2019)Xiao, Shahsafi, Wan, Roney, Joe, Yu, Salman, and Kats}]{Xiao19}
\bibinfo{author}{Y.~Xiao}, \bibinfo{author}{A.~Shahsafi}, \bibinfo{author}{C.~Wan}, \bibinfo{author}{P.~J. Roney}, \bibinfo{author}{G.~Joe}, \bibinfo{author}{Z.~Yu}, \bibinfo{author}{J.~Salman}, \bibinfo{author}{M.~A. Kats},
\newblock \bibinfo{title}{Measuring thermal emission near room temperature using fourier-transform infrared spectroscopy},
\newblock \bibinfo{journal}{Phys. Rev. Appl.} \bibinfo{volume}{11} (\bibinfo{year}{2019}) \bibinfo{pages}{014026}. \URLprefix \url{https://link.aps.org/doi/10.1103/PhysRevApplied.11.014026}. \DOIprefix\doi{10.1103/PhysRevApplied.11.014026}.
\bibitem[{Ozaki et~al.(2021)Ozaki, Christian~Huck, and Engelsen}]{Ozaki}
\bibinfo{author}{Y.~Ozaki}, \bibinfo{author}{S.~T. Christian~Huck}, \bibinfo{author}{S.~B. Engelsen}, \bibinfo{title}{Near-Infrared Spectroscopy}, \bibinfo{publisher}{Springer Singapore}, \bibinfo{year}{2021}. \DOIprefix\doi{https://doi.org/10.1007/978-981-15-8648-4}.
\bibitem[{Regulska et~al.(2005)Regulska, Samsonowicz, Świsłocka, and Lewandowski}]{REGULSKA2005353}
\bibinfo{author}{E.~Regulska}, \bibinfo{author}{M.~Samsonowicz}, \bibinfo{author}{R.~Świsłocka}, \bibinfo{author}{W.~Lewandowski},
\newblock \bibinfo{title}{Vibrational and {NMR} spectra of alkali metal salts of 3-amino-, 3-hydroxy- and 3-halogenobenzoic acids},
\newblock \bibinfo{journal}{Journal of Molecular Structure} \bibinfo{volume}{744-747} (\bibinfo{year}{2005}) \bibinfo{pages}{353--361}. \DOIprefix\doi{https://doi.org/10.1016/j.molstruc.2004.11.061}.

\end{thebibliography}

\end{document}